\documentstyle[prb,tighten,eqsecnum,aps,twocolumn]{revtex}
\begin{document}
\title{Two-spinon dynamic structure factor of the one-dimensional \\
			$S$=1/2 Heisenberg Antiferromagnet}
\author{Michael Karbach ~
		and Gerhard M\"uller}
\address{Department of Physics, The University of Rhode Island, 
		Kingston RI 02881-0817}
\author{A. Hamid Bougourzi}
\address{Institute of Theoretical Physics, SUNY at Stony Brook, Stony Brook, 
		NY 11794}
\date{\today}
\maketitle
%
%
%
%
\begin{abstract}
The exact expression derived by Bougourzi, Couture, and Kacir for the 2-spinon
contribution to the dynamic spin structure factor $S_{zz}(q,\omega)$ of the
one-dimensional $S$=1/2 Heisenberg antiferromagnet at $T=0$ is evaluated for
direct comparison with finite-chain transition rates ($N\leq 28$) and an
approximate analytical result previously inferred from finite-$N$ data, sum
rules, and Bethe-ansatz calculations. The 2-spinon excitations account for
72.89\% of the total intensity in $S_{zz}(q,\omega)$. The singularity structure
of the exact result is determined analytically and its spectral-weight
distribution evaluated numerically over the entire range of the 2-spinon
continuum.  The leading singularities of the frequency-dependent spin
autocorrelation function, static spin structure factor, and $q$-dependent
susceptibility are determined via sum rules.
\end{abstract}
%
\section{Introduction}
%
%
Notwithstanding the fact that Bethe \cite{Bethe31} found the key that solves the
one-dimensional (1D) $S$=1/2 Heisenberg model,
\begin{equation}\label{H}
	H=J\sum_{l=1}^N {\bf S}_l \cdot {\bf S}_{l+1},
\end{equation}
as early as 1931, the emergence of explicit results for various physical
quantities (ground-state energy,\cite{Hulthen38} excitation spectrum,
\cite{CP62} magnetization curve, susceptibility, \cite{Griffiths64,YY66}
thermodynamics \cite{Gaudin71}) was slow at first and then faster since around
1960. Interest in this model began to spread far and wide when the first
compounds with quasi-1D magnetic properties were synthesized and investigated
experimentally.

However, the dynamics of the 1D Heisenberg antiferromagnet, i.e. Hamiltonian
(\ref{H}) with $J>0$, has remained elusive to any rigorous approach during all
those years. An exact result for the dynamic spin structure factor
\begin{equation}\label{Sqw}
	S_{zz}(q,\omega) \equiv \frac{1}{N} \sum_{l,n} e^{iqn}
		\int\limits_{-\infty}^{+\infty} dt e^{i\omega t} 
			\langle S_l^z(t)S_{l+n}^z\rangle,
\end{equation} 
in particular, would have been of great value for the interpretation of a host
of experimental data.\cite{NTCPS91}

Significant progress in the understanding of the $T=0$ dynamics resulted from
the observation\cite{MTBB81} that almost all the spectral weight in
$S_{zz}(q,\omega)$ is carried by a special class of Bethe-ansatz solutions with
excitation energies (in units of $J$ henceforth)
\begin{equation}\label{wmq}
	\omega_m(q) = \pi \sin\frac{q}{2} 
		\cos \left(\frac{q}{2}-\frac{q_m}{2}\right)
\end{equation}
for $N\to \infty, \; 0\leq q\leq \pi, \; 0\leq q_m \leq q$. They form a
two-parameter continuum in the $(q,\omega)$-plane bounded by the branches
\begin{equation}\label{wlwu}
	\omega_L(q) = \frac{\pi}{2} \sin q, \qquad 
	\omega_U(q) = \pi \sin \frac{q}{2}.
\end{equation}
These excitations were later named {\it 2-spinon} states. Their density of
states (after rescaling by a factor $2\pi/N$) is:\cite{MTBB81}
\begin{equation}\label{Dqw}
	D(q,\omega) = \frac{\Theta\biglb(\omega-\omega_L(q)\bigrb)
	          \Theta\biglb(\omega_U(q)-\omega\bigrb)}{
			\sqrt{\omega_U^2(q)-\omega^2}}.
\end{equation}		

The $T=0$ dynamic spin structure factor for a finite system with even $N$ and
periodic boundary conditions can be written in the form
\begin{equation}\label{Sqwgm}
	S_{zz}(q,\omega) = 2\pi \sum_\lambda M_\lambda 
						\delta(\omega-\omega_\lambda),
\end{equation}
where $ M_\lambda =|\langle G | S_q^z|\lambda\rangle|^2$ with
$S_q^z=N^{-1/2}\sum_l e^{iql}S_l^z$ are the transition rates between the singlet
$(S_T=0)$ ground state $|G\rangle$ and the triplet ($S_T=1$) states
$|\lambda\rangle$ with finite-$N$ excitation energies $\omega_\lambda$. Among
them are the $N(N+2)/8$ 2-spinon excitations, which contribute most of the
spectral weight.

The finite-chain analysis of Ref. \onlinecite{MTBB81} suggested that the scaled
transition rates $NM_\lambda$ vary smoothly with $q$ and $\omega$. The
consequence could be that the exact 2-spinon part of the dynamic structure
factor is expressible, in the limit $N\to\infty$, as a product
\begin{equation}\label{SMD}
	 S_{zz}^{(2)}(q,\omega) = M(q,\omega)D(q,\omega),
\end{equation}
with a smooth transition-rate function $M(q,\omega)$, toward which the scaled
finite-$N$ transition rates converge. This scenario is indeed realized in the
related {\it XX} model,\cite{Niemeijer66,KHS70} where the 2-spinon density of
states is given by (\ref{Dqw}) with modified spectral boundaries, and the
transition-rate function is a constant.\cite{MTBB81}
%
\section{Approximate transition rates}
%
In the Heisenberg model (\ref{H}), the finite-$N$ data for the 2-spinon matrix
elements indicate that $M(q,\omega)$ diverges at $\omega=\omega_L(q)$ and
vanishes at $\omega=\omega_U(q)$. In Ref.\onlinecite{MTBB81} the expression
\begin{equation}\label{Mqwgm}
	M^{(a)}(q,\omega) = 
		\sqrt{\frac{\omega_U^2(q)-\omega^2}{\omega^2-\omega_L^2(q)}}
\end{equation}
for the 2-spinon transition-rate function was proposed on the basis of this
observation and the following three requirements: The resulting (approximate)
2-spinon dynamic structure factor
$S_{zz}^{(a)}(q,\omega)=M^{(a)}(q,\omega)D(q,\omega)$ must produce, for $q=\pi$,
the correct infrared exponent.\cite{LP75,FOOTNOTE1} Furthermore, it must produce
the correct $q$-dependence of the known first frequency
moment,\cite{MTBB81,HB74}
\begin{eqnarray}\label{K1}
	K_1(q) &\equiv& \int\limits_0^\infty 
		\frac{d\omega}{2\pi} \omega S_{zz}(q,\omega) 
	    = -\frac{2E_G}{3N} (1-\cos q),
\end{eqnarray}
where $E_G=-N(\ln 2 -1/4)$ is the ground-state energy,\cite{Hulthen38} and, via
the sum rule
\begin{equation}\label{chiq}
	\chi(q) \equiv \frac{1}{\pi} \int\limits_0^\infty  
		\frac{d\omega}{\omega} S_{zz}(q,\omega) ,
\end{equation}
the correct value for the direct susceptibility:\cite{Griffiths64,YY66}
$\chi(0)=1/\pi^2$. The resulting approximate expression,\cite{FOOTNOTE2}
\begin{eqnarray}\label{Szzqwgm}
	S_{zz}^{(a)}(q,\omega) &=& \frac{\Theta\biglb(\omega-\omega_L(q)\bigrb)
   \Theta\biglb(\omega_U(q)-\omega\bigrb)}{\sqrt{\omega^2-\omega_L^2(q)}},
\end{eqnarray}
for the 2-spinon dynamic structure factor has been used quite frequently for the
interpretation of inelastic neutron scattering measurements on a number of
quasi-1D antiferromagnets at low temperature, \cite{NTCPS91} and
for comparisons with the results of various computational
studies.\cite{HRD93,Hallberg95,FKMW95} 

It is interesting to note in this context that the exact dynamic structure
factor $S_{zz}(q,\omega)$ of the Haldane-Shastry model has a structure very
similar to (\ref{Szzqwgm}).\cite{HZ93} In that model, as in the {\it XX} model,
all the spectral weight of $S_{zz}(q,\omega)$ is carried by the 2-spinon
excitations.
%
\section{Exact Transition rates}
%
A detailed assessment of the merits and limitations of the result
(\ref{Szzqwgm}) has become possible only recently through a remarkable new
development.  By approaches based on the concept of infinite-dimensional
symmetries which had been developed in the context of string theory, conformal
field theory, and quantum groups\cite{JM95} Bougourzi, Couture, and Kacir
\cite{BCK96} were able to derive the exact expression for the 2-spinon
transition-rate function in the form\cite{note:factor} 
\begin{equation}\label{Mzqwexact}
	M(q,\omega) = \frac{1}{2} e^{-I(t)}
\end{equation}
where $t=2(\beta_1-\beta_2)/\pi$ and 
\begin{equation}\label{It}
	I(t) = \int\limits_0^\infty dx 
	       \frac{\cosh(2x)\cos (xt) -1}{x\sinh(2x)\cosh x} e^{x}, 
\end{equation}
\begin{mathletters}\label{wqeq}
\begin{eqnarray}
	\omega 
&=&
	\frac{\pi}{2\cosh \beta_1} + \frac{\pi}{2\cosh \beta_2}, \\
	q 
&=& 
	-\cot^{-1}(\sinh \beta_1) - \cot^{-1}(\sinh \beta_2).
\end{eqnarray}
\end{mathletters}
By solving Eqs. (\ref{wqeq}) we can express the auxiliary variable $t$ as a
function of the two physical variables $q,\omega$:
\begin{equation}\label{tqw}
	t=\frac{4}{\pi} {\rm cosh}^{-1} 
		\sqrt{\frac{\omega_U^2(q)-\omega_L^2(q)}{\omega^2-\omega_L^2(q)}}.
\end{equation}
For the numerical evaluation of (\ref{Mzqwexact}) we separate the singular part
from the integral (\ref{It}):
\begin{equation}\label{Itsing}
	I(t) = -I_0 -\ln\left( t\sinh^2 \frac{\pi t}{4} \right) + h(t),
\end{equation}
where
\begin{mathletters}\label{Itnum}
\begin{eqnarray}
	h(t)   &=& {\rm Ci}(t) + f_1(t)-f_2(t), \label{Itnuma} \\
	f_1(t) &=& \int\limits_1^\infty \frac{dx}{x} \frac{\cos(xt)}{\cosh^2 x},
	\quad
	f_2(t) = \int\limits_0^1 \frac{dx}{x} \; \frac{\cos(xt)}{\coth^2 x}, \\
	I_0 &=& \gamma+f_1(0)-f_2(0)=0.3677103....
\end{eqnarray}
\end{mathletters}
A series expansion of $s(t)\equiv[\ln t + C - h(t)]/2$,
\[
	s(t)=\int\limits_0^\infty dx \frac{\sin^2(xt/2)}{x\cosh^2 x} =
	\sum_{m=1}^\infty (-1)^m m \ln \left(1+\frac{t^2}{4m^2}\right),
\]
\begin{equation}
	C \equiv e^{I_0}/2 = 0.72221\ldots ,
\end{equation}
brings (\ref{Mzqwexact}) with $t$ from (\ref{tqw}) into closed form:
\begin{eqnarray}
	M(q,\omega) = 
&& 
	C t \sinh\left(\frac{\pi t}{4}\right)
\nonumber \\ && ~~~
	\times\prod_{m=1}^\infty 
		\frac{\{1+[t/(4m-2)]^2\}^{2m-1}}{\{1+[t/4m]^2\}^{2m}}.
\end{eqnarray}

The exact 2-spinon part of $S_{zz}(q,\omega)$, i.e.  the function
(\ref{SMD}) with the density of states (\ref{Dqw}) and the transition-rate
function (\ref{Mzqwexact}) evaluated numerically via (\ref{Itsing}) with
(\ref{tqw}) is plotted in Fig.  1(a). For comparison, the approximate result
(\ref{Szzqwgm}) is shown in Fig.  1(b). The two results look very similar, yet
there are subtle differences, which may not matter for most experimental
comparisons but are important for comparisons with other theoretical results.

Both expressions diverge at the lower spectral boundary $\omega_L(q)$.  At the
upper boundary $\omega_U(q)$, $S_{zz}^{(a)}(q,\omega)$ has discontinuity,
whereas $S_{zz}^{(2)}(q,\omega)$ approaches zero continuously over a rounded
shoulder. 

The structure of the exact transition rate function (\ref{Mzqwexact}) lends
itself naturally to be factorized into the approximate function (\ref{Mqwgm})
and a correction which accounts for the modified singularities at the boundaries
of the 2-spinon continuum:
\begin{equation}
	M(q,\omega)=M^{(a)}(q,\omega)\sqrt{Ct/2}e^{h(t)/2}.
\end{equation}
%
\section{Singularities at $\omega_L(\lowercase{q})$ and 
			$\omega_U(\lowercase{q})$}
%
What is the precise nature of the leading singularity in the transition-rate
function $M(q,\omega)$ and in the 2-spinon dynamic structure factor
$S_{zz}^{(2)}(q,\omega)$ at the spectral boundaries $\omega_U(q)$ and
$\omega_L(q)$, and how do these singularities compare with those of the
approximate results $S_{zz}^{(a)}(q,\omega)$ and $M^{(a)}(q,\omega)$? The answer
is obtained by inserting (\ref{Itsing}) into (\ref{Mzqwexact}), evaluating the
leading term for $t\to 0$ and $t\to\infty$, respectively, and inserting
(\ref{tqw}) expanded accordingly.

At $\omega_U(q)$ the transition-rate function is thus found to approach zero
linearly,
\begin{equation}
	M(q,\omega) \stackrel{\omega\to\omega_U}{\longrightarrow}
		\frac{8C}{\pi}\frac{\omega_U(q)}{\omega_U^2(q)-\omega_L^2(q)}
		[\omega_U(q)-\omega],
\end{equation}
which implies that the 2-spinon dynamic structure factor vanishes in a
square-root cusp:
\begin{equation}
	S_{zz}^{(2)}(q,\omega) \stackrel{\omega\to\omega_U}{\longrightarrow}
		\frac{8C}{\pi}\frac{\sqrt{2\omega_U(q)}}{\omega_U^2(q)-\omega_L^2(q)}
	\sqrt{\omega_U(q)-\omega}.
\end{equation}
$M^{(a)}(q,\omega)$ vanishes more slowly, $\sim [\omega_U(q)-\omega]^{1/2}$,
implying that $S_{zz}^{(a)}(q,\omega)$ drops to zero abruptly.

At $\omega_L(q)$ we find a square-root divergence (for $q\neq \pi$) in both the
exact and the approximate transition-rate functions, but in the former this
power-law singularity is accompanied by a logarithmic correction:
\begin{eqnarray}\label{Mqwtowl}
	M(q,\omega)  \stackrel{\omega\to\omega_L}{\longrightarrow} 
&&
		\frac{\sqrt{C/2}}{\pi}
		\sqrt{\frac{\omega_U^2(q)-\omega_L^2(q)}{\omega_L(q)}} 
\nonumber \\ && ~~
		\times \frac{1}{\sqrt{\omega-\omega_L(q)}}
			\sqrt{\ln{\frac{1}{\omega-\omega_L(q)}}}.
\end{eqnarray}
Since the 2-spinon density of states is a step function near $\omega_L(q)$, only
the prefactor changes in $S_{zz}^{(2)}(q,\omega)$:
\begin{equation}
	S_{zz}^{(2)}(q,\omega) \stackrel{\omega\to\omega_L}{\longrightarrow}  
		\frac{M(q,\omega)}{\sqrt{\omega^2(q)-\omega_L^2(q)}}.
\end{equation}
For $q\to\pi$ the singularity at $\omega_L(q)$ turns into a much stronger
infrared singularity:
\begin{eqnarray}
	M(\pi,\omega) && \stackrel{\omega\to 0}{\longrightarrow}
		\sqrt{2\pi C} \frac{1}{\omega}\sqrt{\ln\frac{1}{\omega}}, \\
	S_{zz}^{(2)}(\pi,\omega) && \stackrel{\omega\to 0}{\longrightarrow} 
		\sqrt{\frac{2C}{\pi}} 
		\frac{1}{\omega}\sqrt{\ln\frac{1}{\omega}}. \label{Szz2wto0}
\end{eqnarray}
%
\section{Spin autocorrelation function}
%
A quantity of some interest in various experimental and theoretical contexts is
the frequency-dependent spin autocorrelation function
\begin{equation}
	\Phi_{zz}(\omega) \equiv \int\limits_{-\infty}^{+\infty}
				dt\; e^{-i\omega t} 
			\langle S_l^z(t)S_l^z \rangle.
\end{equation}
The 2-spinon contribution to $\Phi_{zz}(\omega)$,
\begin{equation}
	\Phi_{zz}^{(2)}(\omega) \equiv \frac{1}{\pi}\int\limits_0^\pi 
		dq \; S_{zz}^{(2)}(q,\omega),
\end{equation}
is a piecewise smooth function over the range of 2-spinon energies
$0<\omega<\pi$ and has singularities at $\omega=0,\pi/2,\pi $.  The approximate
2-spinon autocorrelation function inferred from (\ref{Szzqwgm}) can be evaluated
in terms of elliptic integrals.  It has a step discontinuity at $\omega=0$,
\begin{equation}
	\Phi_{zz}^{(a)}(\omega) \stackrel{\omega\to 0}{\longrightarrow}
		\frac{1}{\pi } + O(\omega),
\end{equation}
a logarithmic divergence at $\omega=\pi /2$,
\begin{equation}	
	\Phi_{zz}^{(a)}(\omega) 
	\stackrel{\omega\to \pi /2}{\longrightarrow}
		\propto \ln\frac{1}{|\pi /2-\omega|},
\end{equation}
and a square-root cusp at $\omega=\pi $,
\begin{equation}
	\Phi_{zz}^{(a)}(\omega) \stackrel{\omega\to \pi }{\longrightarrow}
		\propto \sqrt{\pi -\omega}.
\end{equation}

The exact 2-spinon expression has logarithmic divergences at $\omega=0,\pi/2$,
and a linear cusp at $\omega=\pi$:
\begin{equation}
	\Phi_{zz}^{(2)}(\omega) \stackrel{\omega\to 0}{\longrightarrow}
	\propto \ln\frac{1}{\omega}.
\end{equation}
\begin{equation}	
	\Phi_{zz}^{(2)}(\omega)  \stackrel{\omega\to \pi /2}{\longrightarrow}
		\propto	\left(\ln\frac{1}{|\pi /2-\omega|}\right)^{3/2},	
\end{equation}
\begin{equation}
	\Phi_{zz}^{(2)}(\omega) \stackrel{\omega\to \pi}{\longrightarrow}
	\propto (\pi  - \omega)
\end{equation}
The functions $\Phi_{zz}^{(2)}(\omega)$ and $\Phi_{zz}^{(a)}(\omega)$ are
plotted in Fig. 2. 	
%
\section{Sum rules}
%
How important is the 2-spinon contribution to $S_{zz}(q,\omega)$ in relation to
that of other excited states? The key to the answer is provided by sum rules,
such as the first frequency moment (\ref{K1}), which is known for all $q$, or
the susceptibility (\ref{chiq}), which is known for $q=0$ only, or the
integrated intensity (static structure factor),
\begin{equation}\label{I}
	I(q) \equiv \int\limits_0^\infty \frac{d\omega}{2\pi} S_{zz}(q,\omega),
\end{equation}
of which we know the grand total:
\begin{equation}
	I_T = \frac{1}{\pi}\int\limits_0^{\pi} dq\; I(q) = 
		\langle (S_l^z)^2 \rangle = \frac{1}{4}.
\end{equation}
The exact 2-spinon contribution to the $n^{th}$ frequency moment of
$S_{zz}(q,\omega)$,
\begin{equation}
	K_n(q) \equiv \int\limits_0^\infty  
		\frac{d\omega}{2\pi} \omega^n S_{zz}(q,\omega),
\end{equation}
as obtained from (\ref{SMD}) with (\ref{Dqw}) and (\ref{Mzqwexact}) can be
brought into the form
\begin{equation}\label{Kn2q}
	K_n^{(2)}(q)=\frac{2C}{\pi^3} [\omega_U(q)]^{n+1} k_n(q),
\end{equation}
where
\begin{equation}\nonumber
	k_n(q)=\int\limits_0^\infty dx \frac{x\sinh x}{\cosh^2 x} 
	\left(1-\sin^2\frac{q}{2} \tanh^2 x \right)^{\frac{n-1}{2}} e^{-s(4x/\pi)}.
\end{equation}
For $n=2m+1=1,3,...$ this expression reduces to a polynomial in $\cos q$,
\begin{mathletters}
\begin{eqnarray}
	K_{2m+1}^{(2)}(q) 
&& = \nonumber \\ && \hspace*{-18mm}
		\frac{C}{\pi}\left(\frac{\pi^2}{2}\right)^m 
			\sum_{l=0}^m \left(\begin{array}{c} m \\ l \end{array} \right)
					\frac{(-1)^l}{2^l}\kappa_l(1-\cos q)^{m+1+l}, \\
	\kappa_l 
&\equiv& 
	\int\limits_0^\infty dx \frac{x(\tanh x)^{2l+1}}{\cosh x} e^{-s(4x/\pi)}.
\end{eqnarray}
\end{mathletters}
The exact sum rules for $K_{2m+1}(q)$ were shown to have precisely this general
structure,\cite{FKM96,Mueller82} which, incidentally, is also reproduced by the
frequency moments $K_{2m+1}^{(a)}(q)$ of $S_{zz}^{(a)}(q,\omega)$. However, the
exact coefficients of the polynomial are only known for $m=0$. Comparison of
\begin{equation}
	K_1^{(2)}(q) = \frac{C}{\pi} \kappa_0(1-\cos q), \quad \kappa_0=0.9163...,
\end{equation}
with (\ref{K1}) provides one way of measuring the relative spectral weight of
the 2-spinon excitations:
\begin{equation}
	\frac{K_1^{(2)}(q)}{K_1(q)} = 0.7130...
\end{equation}
A somewhat larger share of spectral weight, $K_1^{(a)}(q)/$
$K_1(q)\!=\!0.8462...$, is accounted for by $S_{zz}^{(a)}(q,\omega)$.

A different way of measuring the relative 2-spinon spectral weight is provided
by the static structure factor (\ref{I}). Here, the missing spectral weight of
higher-lying excitations is weighted less heavily.

The exact 2-spinon static structure factor $I^{(2)}(q)=K_0^{(2)}(q)$ taken from
(\ref{Kn2q}) and integrated over $q$ yields the total 2-spinon intensity
\begin{equation}
	I_T^{(2)} = \frac{4C}{\pi^3}\int\limits_0^\infty dx
		\frac{x^2}{\cosh x} e^{-s(4x/\pi)}  \simeq 0.7289 I_T.
\end{equation}
The total intensity of $S_{zz}^{(a)}(q,\omega)$ is \cite{MTBB81}
$I_T^{(a)}\simeq0.7424 I_T$.

The observation that $S_{zz}^{(a)}(q,\omega)$ overestimates the total 2-spinon
intensity by a smaller fraction, $I_T^{(a)}/I_T^{(2)}\simeq 1.0185$, than the
first frequency moment of the 2-spinon spectral weight,
$K_1^{(a)}(q)/K_1^{(2)}(q)\simeq 1.1868$, is consistent with the observation
made previously that it predicts too much spectral weight near $\omega_U(q)$ and
too little near $\omega_L(q)$.

At small $q$, where the 2-spinon continuum is very narrow, all frequency moments
of $S_{zz}^{(2)}(q,\omega)$ and $S_{zz}^{(a)}(q,\omega)$ have exactly the same
ratio
\begin{equation}
	\frac{K_n^{(a)}(q)}{K_n^{(2)}(q)} \stackrel{q\to 0}{\longrightarrow}
		\frac{4C}{\pi}\kappa_0=0.8426...
\end{equation}
The implications of the frequency moments $K_0^{(2)}(q)$ and $K_{-1}^{(2)}(q)$
for the singularities of the static structure factor and the static
susceptibility, respectively, will be discussed later.

%
\section{Finite-chain results}
%
To what extent and accuracy can the spectral-weight distribution of
$S_{zz}(q,\omega)$ be reconstructed from (\ref{Sqwgm}) on the basis of
finite-chain data for excitation energies $\omega_\lambda$ and transition rates
$M_\lambda$?  In a generic situation, the chances for success may be remote.
Convergence of the finite-$N$ data for (\ref{Sqwgm}) toward the infinite-$N$
spectral density may only exist in an average sense, such as can be realized, at
least in principle, by a histogram representation of (\ref{Sqwgm}), but hardly
in practice given the very coarse-grained spectral-weight distribution even in
the largest systems that can be handled computationally.

Among the ever growing collection of Bethe-ansatz solvable models, there exist
numerous situations where the spectral density of interest is dominated by a
specific class of excitations that can be identified in terms of Bethe quantum
numbers.  When the dynamically dominant class of excitations consists of a
two-parameter continuum, as is frequently the case, the task of reconstructing
that spectral density from finite-$N$ data with reasonable accuracy may be
perfectly within the reach of state-of-the-art computational applications.

In the case at hand, the 2-spinon excitation energies $\omega_\lambda$ can be
evaluated for finite chains over a wide range of $N$ and then again for infinite
$N$, all via Bethe ansatz. The finite-$N$ transition rates $M_\lambda$ can be
evaluated directly from the Bethe-ansatz wave function for the ground state and
the 2-spinon states up to $N=16$ and indirectly from the finite-$N$ ground-state
wave function via the recursion method \cite{FKMW95} up to $N=28$.

The crucial point for the reconstruction of the 2-spinon part of the dynamic
structure factor $S_{zz}(q,\omega)$ is that it factorizes into two smooth
functions: the density of states $D(q,\omega)$, which can be determined exactly
via Bethe ansatz, and the transition rate function $M(q,\omega)$, toward which
the finite-$N$ transition rates seem to converge in the following sense: pick
any sequence of finite-$N$ 2-spinon states with energies $\omega_\lambda(N)$ and
wave numbers $q_\lambda(N)$ converging toward $(q,\omega)$ as $N\to\infty$. Then
the associated scaled transition rates $NM_\lambda(N)$ converge toward the exact
transition rate function $M(q,\omega)$.

In the main plot of Fig. 3 we show the transition rate functions $M(\pi,\omega)$
(exact, solid line) and $M^{(a)}(\pi,\omega)$ (approximate, dashed line) along
with scaled finite-$N$ transition rates $NM_\lambda$ for $N=6,8,...,28$. The
downward deviation of $M^{(a)}(\pi,\omega)$ from $M(\pi,\omega)$ at low
frequencies is due to the lacking logarithmic corrections in the infrared
divergence and the upward deviation at high frequencies due to the different
cusp singularity at $\omega_U(\pi)$.

All finite-$N$ data points fall close to the solid line. Their deviations from
that line have an irregular appearance at first sight. This is attributable to
the fact that an increasing number of spectral contributions from systems with
increasing $N$ are distributed over a fixed frequency interval. However, when we
focus on the lowest-lying excitation, for example, we see that the data points
move away from the dashed line toward the solid line. The uniform convergence of
this particular sequence of data points is best observable in the representation
of the inset on the left of Fig. 3. 

The region near $\omega_U(\pi)$ is shown magnified in the inset on the
right. Here the finite-$N$ data converge in a much more complicated
pattern. Nevertheless, the trend is clearly toward the linear behavior of the
solid line and away from the square-root behavior of the dashed line.  

The corresponding results for $q=\pi/2$ are depicted in Fig. 4. Here the highest
2-spinon excitation for $N=28$, which we were unable to compute with sufficient
accuracy via the recursion method, is not included. Even with the few
finite-chain data points available in this case, the finite-size scaling
behavior of the transition rates $M_\lambda$ and their convergence toward the
exact transition-rate function is again convincingly determined.

Given the exact asymptotic finite-size gap of the lowest 2-spinon excitation at
$q=\pi$,\cite{WE87}
\begin{equation}\label{fssgap}
	\omega_1 \stackrel{N\to\infty}{\longrightarrow}\frac{\alpha}{N},
	\quad \alpha=\frac{\pi^2}{2},
\end{equation}
and the exact infrared divergence (\ref{Szz2wto0}) of $S_{zz}^{(2)}(q,\omega)$,
it is possible to determine, under standard scaling assumptions, the leading
$N$-dependence of the integrated intensity at $q=\pi$,
\begin{equation}
	I(\pi,N) \stackrel{N\to\infty}{\longrightarrow} 
	\frac{m_0}{2\pi}(\ln N)^{3/2}
\end{equation}
with $m_0=\sqrt{2C/\pi}$. The exact coefficient, $m_0/2\pi=0.1079...$, is
significantly higher than the value 0.09052 predicted in a recent DMRG
study.\cite{HHM95} The leading singularity of the integrated intensity for
$N=\infty,\; q\to\pi$ is then predicted to be of the form
\begin{equation}
	I(q,\infty) \stackrel{q\to\pi}{\longrightarrow}
		\frac{m_0}{2\pi} \left[-\ln\left(1-\frac{q}{\pi}\right)\right]^{3/2},
\end{equation}
which is consistend with the exactly known leading asymptotic term of the static
spin correlation function\cite{SFS89} $\langle S_l^zS_{l+n}^z\rangle \sim
(-1)^nn^{-1}(\ln n)^{1/2}/n$. 

The corresponding leading terms for the static susceptibility read:
\begin{eqnarray}
	\chi(\pi,N)  &\stackrel{N\to\infty}{\longrightarrow}& 
		\frac{m_0}{\pi\alpha}N\sqrt{N}, \\
	\chi(q,\infty)  &\stackrel{q\to\pi}{\longrightarrow}& \propto 
		\frac{\sqrt{-\ln(\pi-q)}}{\pi-q}.
\end{eqnarray}
%
\acknowledgements
%
The work at URI was supported by NSF Grant DMR-93-12252, and by the Max Kade
Foundation. The work at SUNYSB was supported by NSF Grant PHY-93-09888.
A.H.B. would like to thank M. Couture for encouragements and stimulating
discussions. Access to the computing facilities at the national Center for
Supercomputing Applications, University of Illinois at Urbana-Champaign is
gratefully acknowledged.
%
%

%
%
\newpage
\begin{figure}\label{fig_stf}
	\caption{(a) Exact and (b) approximate 2-spinon dynamic structure
	factor. Both expressions are nonzero only in the shaded region of the
	$(q,\omega)$-plane bounded by $\omega_L(q)$ and $\omega_U(q)$.}
\end{figure}
\begin{figure}\label{fig_autocorrelation}
	\caption{Two-spinon part of the frequency-dependent spin autocorrelation
	function. The solid line represents the exact result
	$\Phi_{zz}^{(2)}(\omega)$ and the dashed line the approximate result
	$\Phi_{zz}^{a}(\omega)$} .
\end{figure}
\begin{figure}\label{fig_mzqpiw}
	\caption{Two-spinon transition-rate function at $q=\pi$. The solid line
	represents the exact result $M(q,\omega)$ and the dashed line the
	approximate result $M^{(a)}(q,\omega)$. Also shown are scaled finite-chain
	transition rates $NM_\lambda$ for all 2-spinon excitations at $q=\pi$ of
	systems with $N=6,8,...,16,28$ spins, and for lowest 2-spinon excitations
	also of systems with $N=18,20,...,26$.  The low-frequency and high-frequency
	parts are shown again in the insets with transformed scales on both axes.}
\end{figure}
\begin{figure}\label{fig_mzqp12w}
	\caption{Two-spinon transition-rate function at $q=\pi/2$. The solid line
	represents the exact result $M(q,\omega)$ and the dashed line the
	approximate result $M^{(a)}(q,\omega)$. Also shown are scaled finite-chain
	transition rates $NM_\lambda$ for all 2-spinon excitations at $q=\pi$ of
	systems with $N=8,12,16,28$ spins.  The low-frequency and high-frequency
	parts are shown again in the insets with transformed scales on both axes.}
\end{figure}

\begin{thebibliography}{10}

\bibitem{Bethe31}
	H. Bethe, 
	Z. Phys. {\bf 71},  205  (1931).

\bibitem{Hulthen38}
	L. Hulth\'en, 
		Arkiv Mat. Astron. Fysik A11 {\bf 26},  1  (1938).

\bibitem{CP62}
	{J. des Cloizeaux} and {{J.J.} Pearson}, 
		Phys. Rev. {\bf 128},  2131  (1962).

\bibitem{Griffiths64}
	{{R.B.} Griffiths}, 
		Phys. Rev. {\bf 133}, A768  (1964).

\bibitem{YY66}
	{{C.N.} Yang} and {{C.P.} Yang}, 
		Phys. Rev. {\bf 150},  321  (1966); 
				   {\bf 150},  327  (1966);
            	   {\bf 151},  258  (1966).

\bibitem{Gaudin71}
	M. Gaudin, 
		Phys. Rev. Lett. {\bf 26},  1301  (1971);
	M. Takahashi, 
		Prog. Theor. Phys. {\bf 46},  401  (1971).

\bibitem{NTCPS91}
	{S.E.} Nagler {\it et al.}, 
		Phys. Rev. B {\bf 44},  12361  (1991);
	{{D.A.} Tennant}, {{T.G.} Perring}, {{R.A.} Cowley}, and {{S.E.} Nagler}, 
		Phys. Rev. Lett. {\bf 70},  4003  (1993);
	{D.C.} Dender {\it et al.}, 
		Phys. Rev. B {\bf 53}, 2583 (1996).

\bibitem{MTBB81}
	{G. M\"uller}, {H. Thomas}, {H. Beck}, and {{J.C.} Bonner}, 
		Phys. Rev. B {\bf 24},  1429  (1981).

\bibitem{Niemeijer66}
	T. Niemeijer, 
		Physica {\bf 36},  377  (1966).

\bibitem{KHS70}
	{S. Katsura}, {T. Horiguchi}, and {M. Suzuki}, 
		Physica {\bf 46},  67  (1970).

\bibitem{LP75}
	{A. Luther} and {I. Peschel}, 
		Phys. Rev. B {\bf 12},  3908  (1975).

\bibitem{FOOTNOTE1}
The logarithmic corrections to the $\sim \omega^{-1}$ singularity were not
known at the time.

\bibitem{HB74}
	{{P.C.} Hohenberg} and {{W.F.} Brinkman}, 
		Phys. Rev. B {\bf 10},  128  (1974).

\bibitem{FOOTNOTE2}
Alternative requirements considered in Ref. 8 yield prefactors in (8) which are
  slightly greater than one.

\bibitem{HRD93}
	{S. Haas}, {J. Riera}, and {E. Dagotto}, 
		Phys. Rev. B {\bf 48},  3281  (1993).

\bibitem{Hallberg95}
	K. Hallberg, 
		Phys. Rev. B {\bf 52},  R9827  (1995).

\bibitem{FKMW95}
	{A. Fledderjohann}, {M. Karbach}, {{K.-H.} M\"utter}, and {P. Wielath}, 
		J. Phys.: Condens. Matter {\bf 7},  8993  (1995).

\bibitem{HZ93}
	{{F.D.M.} Haldane} and {{M.R.} Zirnbauer}, 
		Phys. Rev. Lett. {\bf 71},  4055 (1993).

\bibitem{JM95}
		{M. Jimbo} and  {T. Miwa},
		{\it Algebraic Analysis of Solvable Lattice Models},
		(American Mathematical Society, CBMS, 1995).

\bibitem{BCK96}
	{{A.H.} Bougourzi}, {M. Couture}, and {M. Kacir}, 
		Preprint ITP-SB-96-21, Stony Brook.

\bibitem{note:factor}	We have found and corrected a discrepancy in chapter
		10.4 of Ref.~\onlinecite{JM95}, which affects the constant prefactor in
		(\ref{Mzqwexact}). 
	
\bibitem{FKM96}
	{A. Fledderjohann}, {M. Karbach}, and {{K.-H.} M\"utter}, 
		Phys. Rev. B {\bf  53},  11543  (1996).

\bibitem{Mueller82}
	G. M\"uller, 
		Phys. Rev. B {\bf 26},  1311  (1982).

\bibitem{WE87}
	{F. Woynarovich} and {{H.-P.} Eckle}, 
		J. Phys. A: Math. Gen. {\bf 20},  L97 (1987).

\bibitem{HHM95}
	{K. Hallberg}, {P. Horsch}, and {G. Mart\'inez}, 
		Phys. Rev. B {\bf 52},  R719 (1995).

\bibitem{SFS89}
	{{R.R.P.} Singh}, {{M.E.} Fisher}, and {R. Shankar}, 
		Phys. Rev. B {\bf 39},  2562 (1989).

\end{thebibliography}
\end{document}